\documentclass[11pt]{article}
\usepackage{epsfig}
\usepackage{oldgerm}
\input{amssym}
\def\be{\begin{equation}}
\def\ee{\end{equation}}

\newsavebox{\PSLASH}
\sbox{\PSLASH}{$p$\hspace{-1.8mm}/}

\begin{document}
\title{Patterned and Disordered Continuous Abelian Sandpile Model}
\author{N. Azimi-Tafreshi\footnote{e-mail: azimi@physics.sharif.ir}, S. Moghimi-Araghi\footnote{e-mail: samanimi@sharif.edu}\\
Department of Physics, Sharif University of Technology,\\ Tehran,
P.O.Box: 11365-9161, Iran} \date{} \maketitle

\begin{abstract}
We study critical properties of the continuous Abelian sandpile
model with anisotropies in toppling rules that produce ordered
patterns on it. Also we consider the continuous directed sandpile
model perturbed by a weak quenched randomness and study critical
behavior of the model using perturbative conformal field theory
and show the model has a new random fixed point.
\vspace{5mm}%
\\
\textit{PACS}: 05.65+b, 89.75.Da
\newline \textit{Keywords}: Self-Organized Criticality, Sandpile models, Conformal field theory.
\end{abstract}
\section{Introduction}
The idea of self-organized criticality, introduced by Bak, Tang
and Wisenberg \cite{BTW}, provides a useful framework for the study
of non-equilibrium systems which dynamically evolve into a
critical state without tuning of a control parameter. At critical
state, these systems show scaling behaviors and this scaling behavior of
the system is characterized by critical exponents \cite{Jensen}.

The BTW sandpile model, renamed Abelian sandpile model after
Dhar's work \cite{Dhar}, is the simplest lattice model that
displays self-organized critical behavior. The Abelian
structure of the model allows the theoretical determination of
many of its properties \cite{Prizh, Ivashkevich94}. This model is usually
defined on a square lattice. At each site of the lattice an
integer height variable between $1$ to $4$ is assigned which
represents the number of sand grains of that site. The evolution of the
model at each time step is simple: a grain of sand is added to a
random site. If the height of that site becomes greater than the
critical height $h_{c}=4$, the site will be unstable; it topples and four grains leave the site and each of the four neighbors gets one of the grains.
 As a result, some of the
neighbors may become unstable and toppling continue. The process continues untill no unstable site remains and the avalanche ends. To achieve this, one should let some grains of sand leave the system, and this happens at the boundary sites. Every avalanche can be represented as a sequence of
waves of the topplings such that each site at a wave topples only once
\cite{wave}. While the scaling behavior of avalanches is complex and usually not governed by simple scaling laws, it has been shown that the probability distributions for waves display clear power-law
asymptotic behavior \cite{wave1}. 

The scaling exponents of the system shows little dependence on parameters such as the number of neighbors, however if we make the toppling rule anisotropic, then new universality classes may emerge: selecting a particular transport direction in BTW model, Hwa and Kardar \cite{Hwa} defined an anisotropic sandpile
model such that the grains are allowed to leave the system only at
one edge of the system. They determined the critical exponents
with a dynamical renormalization group method. Dhar and Ramaswamy
\cite{DR} defined a directed version of the BTW model and
determined the critical exponents and the two-point correlation
functions exactly in any dimensions. In \cite{anisotropy}, the
effect of anisotropy in a continuous version of sandpile model
(Zhang model) is investigated. In this paper, a $d$-dimensional
lattice is considered. This $d$-dimensional space is divided to two $a$-dimensional and ($d-a$)-dimensional
subspaces. It is assumed that the energy (sand) is
propagated differently for the two subspaces, but inside the
subspaces the propagation of energy is isotropic. It is then
shown that the peaked energy distribution and critical exponents
of the distribution avalanche sizes are affected by the
anisotropy. In \cite{asymmetry}  two
variations of continuous abelian sandpile model are introduced, the directed model
and the elliptical model. It is shown that the elliptical anisotropy
does not change the universality class of the isotropic model
whereas the critical exponents are sensitive to the directed
anisotropy. Karmakar showed that in a quenched disorder sandpile model, the
symmetric or asymmetric flows of sands in each bond determines the
universality class of the undirected model \cite{quenched}. Also,
a quenched disorder directed sandpile model has the same critical
exponents with the BTW model when the local flow balance exist
between inflow and outflow of sands at a site. Otherwise the model
falls in the universality class of the Manna sandpile model
\cite{pan}. 

The original isotope model could be represented with a conformal field theory known as $c=-2$ theory \cite{c2}. When we insert anisotropy in toppling rules the
rotational symmetry of the lattice is broken and the field theory associated with the model could not be conformal field theory. However, it may be possible to restore the rotational symmetry in
large scale or statistically in an anisotropic sandpile model.  
To do this, one can introduce models that the toppling rules have some
patterns on the lattice in a way that in larger scales there will be no preferred directions; that is, locally you have preferred directions which differ site to site in a regular pattern such that on larger scales the system look isotrope.  Another possibility is to assume a quenched randomness for
anisotropy in toppling of lattice sites; that is, we add
anisotropy to the toppling rule of each site, however the amount
of anisotropy and the preferred direction of anisotropy differs site to site randomly. In this way there may be no preferred
direction statistically.

The question we address in this paper is whether the universality class of these
modified models is different from the original sandpile model or not. We show that in some patterned sandpile models the
universality class is the same as the isotropic Abelian sandpile
model's universality class. However it turns out that the presence of disorder in a
sandpile model may change the universality class of the system. This is done exploiting the replica technique; we consider this anisotropy as a perturbation to the original conformal field theory and use renormalization group to describe the perturbative behaviors of the system \cite {perturb1,perturb}
 
The plan of the paper is as follows: in next section we insert
some anisotropies in the redistribution of sands such that create
some ordered patterns. We obtain the free energy function for
these models with using one to one correspondence between the
recurrent configurations of ASM and the spanning tree
configurations on the same lattice \cite{MajDhar}. The effect of
these types of anisotropies on the critical behaviors of the
system Theoretically and numerically is investigated. Next we consider a position
dependent randomness in the toppling rule. Our procedure is based on the perturbative
renormalization group approach around the conformal field theory
describing the isotropic model and obtain the renormalization
group equations for coupling constants.

\section{Patterned Continuous Sandpile Models }
It is known that the universality class of directed sandpile model is different from the ordinary ASM's \cite{DR,anisotropy}. In the directed model, the sand grains are always drifted toward preferred direction, say up-right corner. 
We would like to see if the directedness is introduced to the model only in small scales, is the universality class changed or not. To this end we add the directedness locally in a way that on
average there will be no preferred direction towards which the
sand grains move.

Consider the continuous ASM on a square lattice composed of $N$
lattice sites \cite{GLJ,CASM}. To each site, a continuous height variable in the
[0,4) interval is assigned. We divide the sites into two groups
$A$ and $B$, such that neighbors of one site in group $A$ belong
to group $B$ and vice versa. We impose an anisotropic toppling
rules for points of these two sublattice differently: when a toppling occurs in an $A$-site $1+\epsilon$ amount of sand is transferred to each
of the right and up neighbors and $1-\epsilon$ amount to the down
and left neighbor sites. In the case that a $B$ site  topples, $1+\epsilon$ amount of sand is given to each of
the left and down sites and $1-\epsilon$ amount of sand is transfered to
the right and up neighbors. Here, $\epsilon$ is a positive real
parameter less than $1$ that controls the amount anisotropy. For
$\epsilon=0$ we will have the isotropic model and $\epsilon=1$
characterizes the fully anisotropic model. This toppling rule means that the $A$ sites try to direct the avalanche towards up-left corner
and the $B$ sites try to direct the avalanche to down-right corner,
thus on average the sands do not move in any specific direction.
In Fig. 1 such a lattice is sketched. If a toppling occurs, the amount of sand transfered via thick lines is $1+\epsilon$ and the amount of sand transfered via thin lines is $1-\epsilon$. It is clear that for $\epsilon=1$ the sands are only
allowed to move along one of the thick zigzag paths and therefore the system becomes essentially a set of one-dimensional sandpile models.  
\begin{figure}[t]
\centering
\epsfig{file=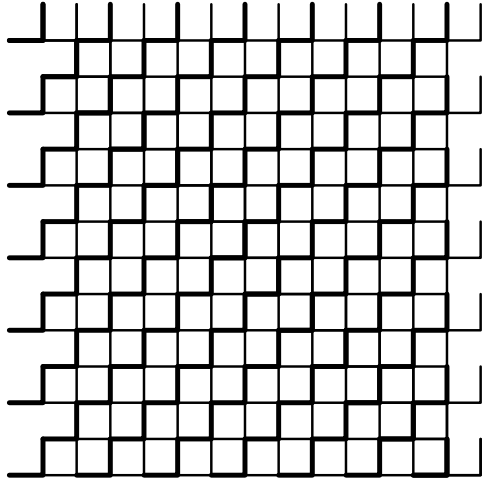, width=.4\linewidth} \caption{patterned
ASM} \label{fig:}
\end{figure}
The elements of the toppling matrix can be written in the following form:
\begin{eqnarray}
\Delta^{A}_{ij,i^{'}j^{'}}=  \left\{
\begin{array}{cc}
4 & i=i^{'},j=j^{'}\\
-(1\pm\epsilon) &  i=i^{'}\pm1\\
-(1\mp\epsilon) &  j=j^{'}\pm1\\
0 & {\rm otherwise}
\end{array}
\right.
\end{eqnarray}
\begin{eqnarray}
\Delta^{B}_{ij,i^{'}j^{'}}=  \left\{
\begin{array}{cc}
4 & i=i^{'},j=j^{'}\\
-(1\mp\epsilon) &  i=i^{'}\pm1\\
-(1\pm\epsilon) &  j=j^{'}\pm1\\
0 & {\rm otherwise}
\end{array}
\right.
\end{eqnarray}

\begin{figure}[t]
\centering
\begin{picture}(100,110)(0,0)
\includegraphics{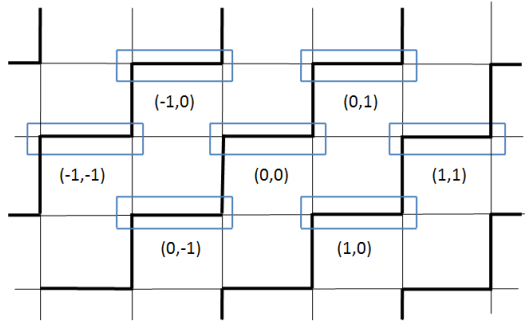}
\end{picture}
\caption{Unit cells of the patterned ASM} \label{fig:}
\end{figure}

A first step to deduce the critical behavior of the system could be finding the free energy function. The closed form of the
free energy is obtained by enumerating 
the corresponding spanning trees on the lattice. The formulation
for enumerating spanning trees for general lattices is given in
\cite{spanning}. We take the unit cells of two lattice sites as
shown in Fig 2. Following the standard procedure, we obtain the
free energy:
\begin{equation}\label{freeenergy}
f=\frac{1}{8\pi^{2}}\int_{0}^{2\pi}d\theta\int_{0}^{2\pi}d\phi\ln\det
F(\theta,\phi)
\end{equation}
where,
\begin{eqnarray}\label{F}
F(\theta,\phi)=& {}&\hspace{-5mm}4I-\left (
a(0,0)+a(1,0)e^{i\theta}+a(-1,0)e^{-i\theta}+a(0,1)e^{i\phi}\right. \nonumber\\
&&\hspace{-7mm}\left.+ a(0,-1)e^{-i\phi}+a(1,1)e^{i(\theta+\phi)}+a(-1,-1)e^{-i(\theta+\phi)}\right)
\end{eqnarray}
and $a(n,\acute{n})$ are the $2\times 2$ cell adjacency matrices
describing the connectivity between sites of the unit cells
$n,\acute{n}$.
\begin{eqnarray}
\label{adjacency}
a(0,0)=\left(
\begin{array}{cc}
  0& 1+\epsilon \\
  1+\epsilon & 0\\
\end{array}
\right),~~~a(0,1)=a^{T}(0,-1)=\left(
\begin{array}{cc}
  0& 0 \\
  1+\epsilon & 0 \\
\end{array}
\right),\nonumber\\
a(-1,0)=a(-1,-1)=a^{T}(1,0)=a^{T}(1,1)=\left(
\begin{array}{cc}
  0& 1-\epsilon \\
  0 & 0 \\
\end{array}
\right)
\end{eqnarray}\\
With a straightforward calculation one finds
\begin{equation}\label{freeenergy2}
f=\frac{1}{8\pi^{2}}\int_{0}^{2\pi}d\theta\int_{0}^{2\pi}d\phi\ln
( 12-4\epsilon^{2}-4(1-\epsilon^{2})\cos \theta
-4(1+\epsilon^{2})\cos \phi\nonumber\\
-2(1-\epsilon^{2})\cos(\theta+\phi)-2(1-\epsilon^{2})\cos(\theta-\phi))
\end{equation}
Now it is seen that the model is equivalent to a free fermion
$8$-vertex model with weights $\{w(1), \ldots w(8) \}$
\cite{freeenergy} that are related to $\epsilon$ with the
following relations:
\begin{eqnarray*}\label{weights}
12-4\epsilon^{2}&=&w(1)^{2}+w(2)^{2}+w(3)^{2}+w(4)^{2}\\
2(1-\epsilon^{2})&=&w(2)w(4)-w(1)w(3)\\
(1-\epsilon^{2})&=&w(5)w(6)-w(3)w(4)\\
2(1+\epsilon^{2})&=&w(2)w(3)-w(1)w(4)\\
0&=&w(5)w(6)-w(7)w(8)
\end{eqnarray*}
The critical properties of the free fermion model are well known
\cite{freeenergy}. It is found that for all values of $\epsilon$
the free energy function is analytical and the model shows no
phase transition. It means although by inserting this kind of
anisotropy some symmetries of the lattice are broken, but the
broken symmetry operator is irrelevant and takes the system to the
original critical fixed point. This fact can be checked by
numerical simulations. We have simulated the model on a square lattice
with sizes $L=64,128,256$ and $512$. After the system arrives at recurrent configurations, we began to collect data. At each size $10^6$ avalanches have been considered to derive the wave statistics.  Fig 3 displays the wave toppling
distributions for different system sizes and three different
values of $\epsilon$. A power law fit to these curves determines
the critical exponent $\tau_{s}^{(w)}$ defined as
$P_{s}^{w}(s)\sim s^{-\tau_{s}^{(w)}}$. In Fig. 4 the extrapolated
value of $\tau$ for $L\rightarrow \infty$ is obtained:
$\tau(\infty)=1.00\pm 0.01$ for $\epsilon=0.1$,
$\tau(\infty)=0.99\pm 0.01$ for $\epsilon=0.4$ and
$\tau(\infty)=1.01\pm 0.01$ for $\epsilon=0.8$. As we see, the wave
exponents are independent of  $\epsilon$ 
and are consistent with the exact value of $\tau_{s}^{(w)}=1$ for
$\epsilon=0$ \cite{waveexponent}.\\
\begin{figure}[b]
\begin{picture}(200,130)(0,0)
\includegraphics{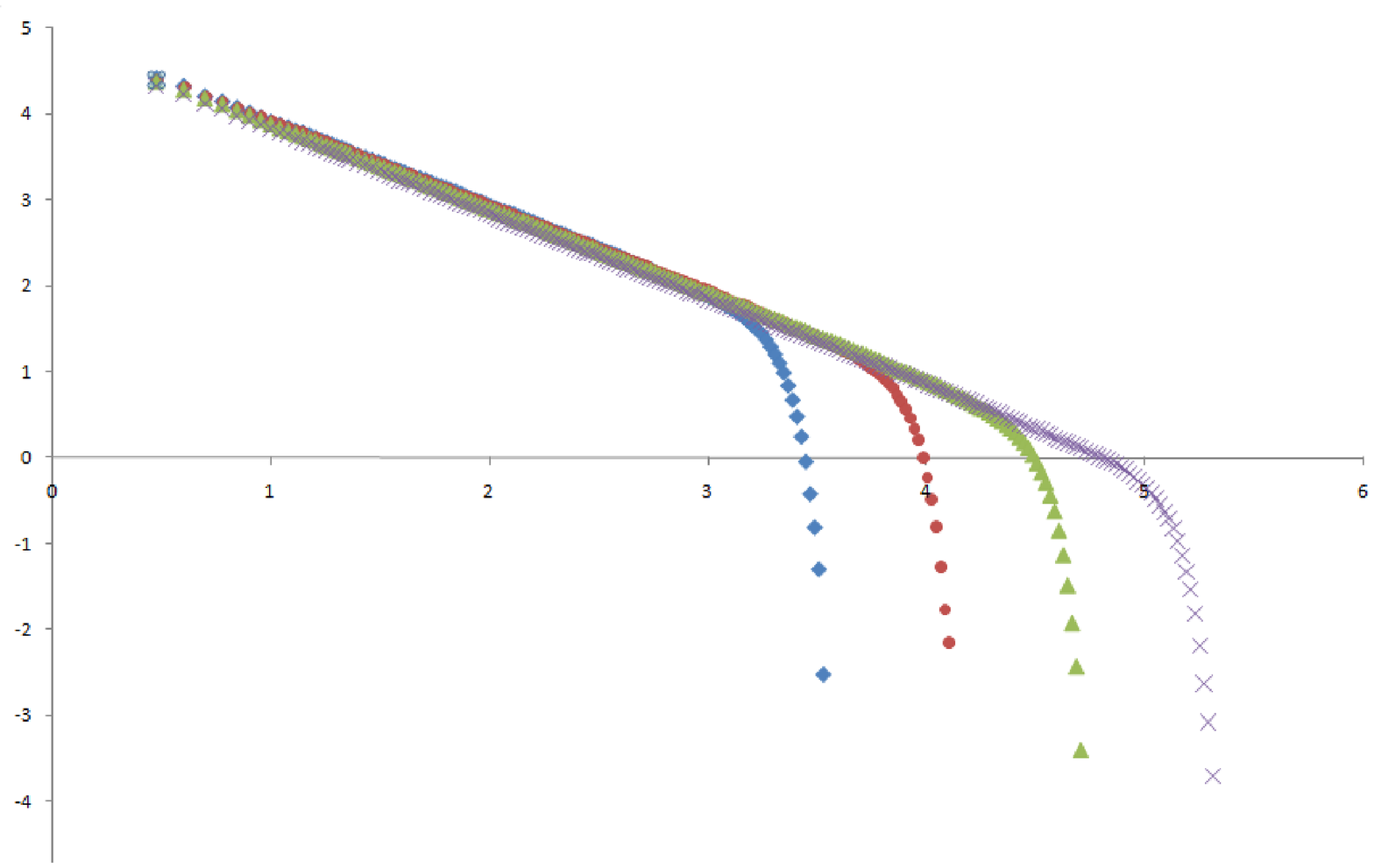} \includegraphics{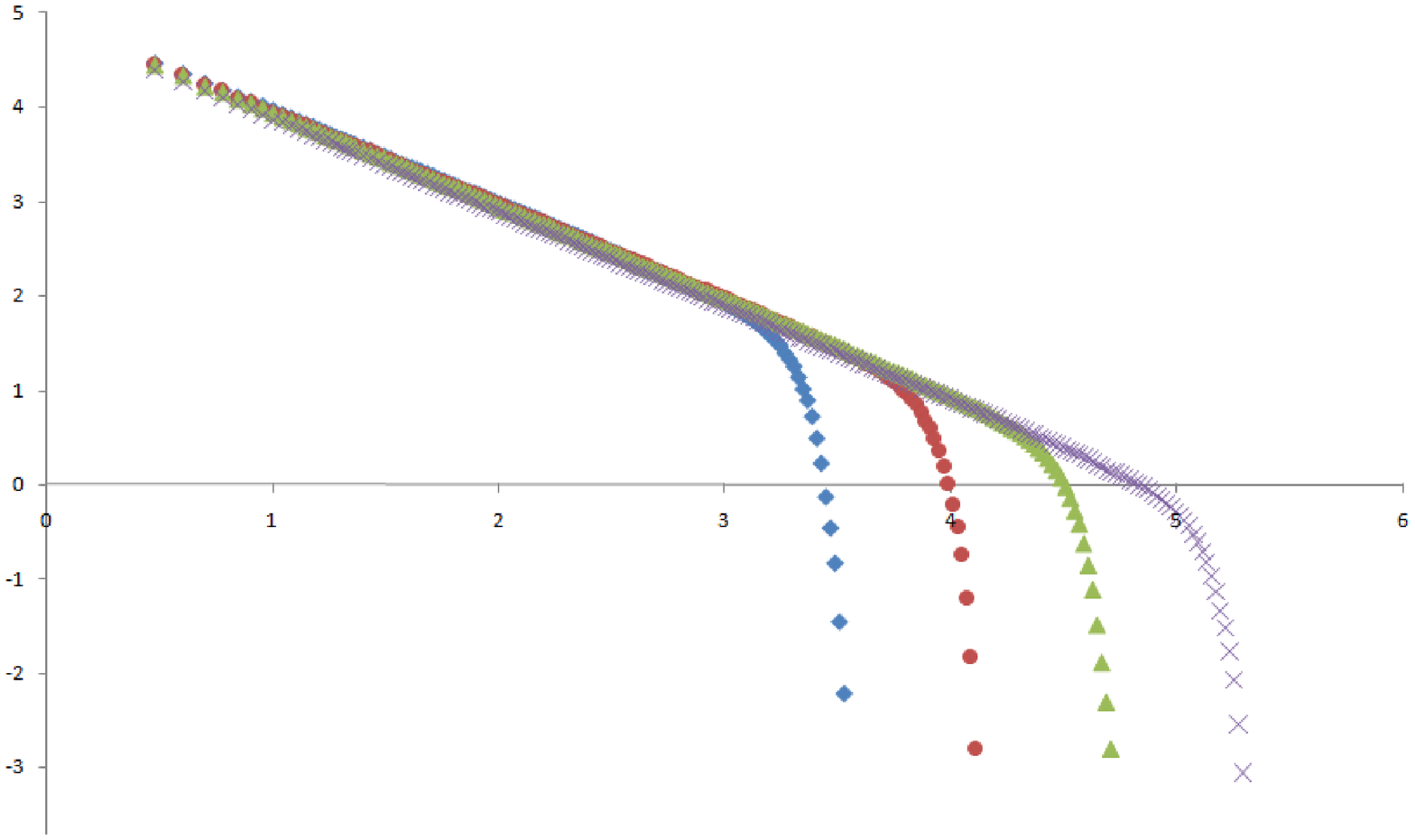} \includegraphics{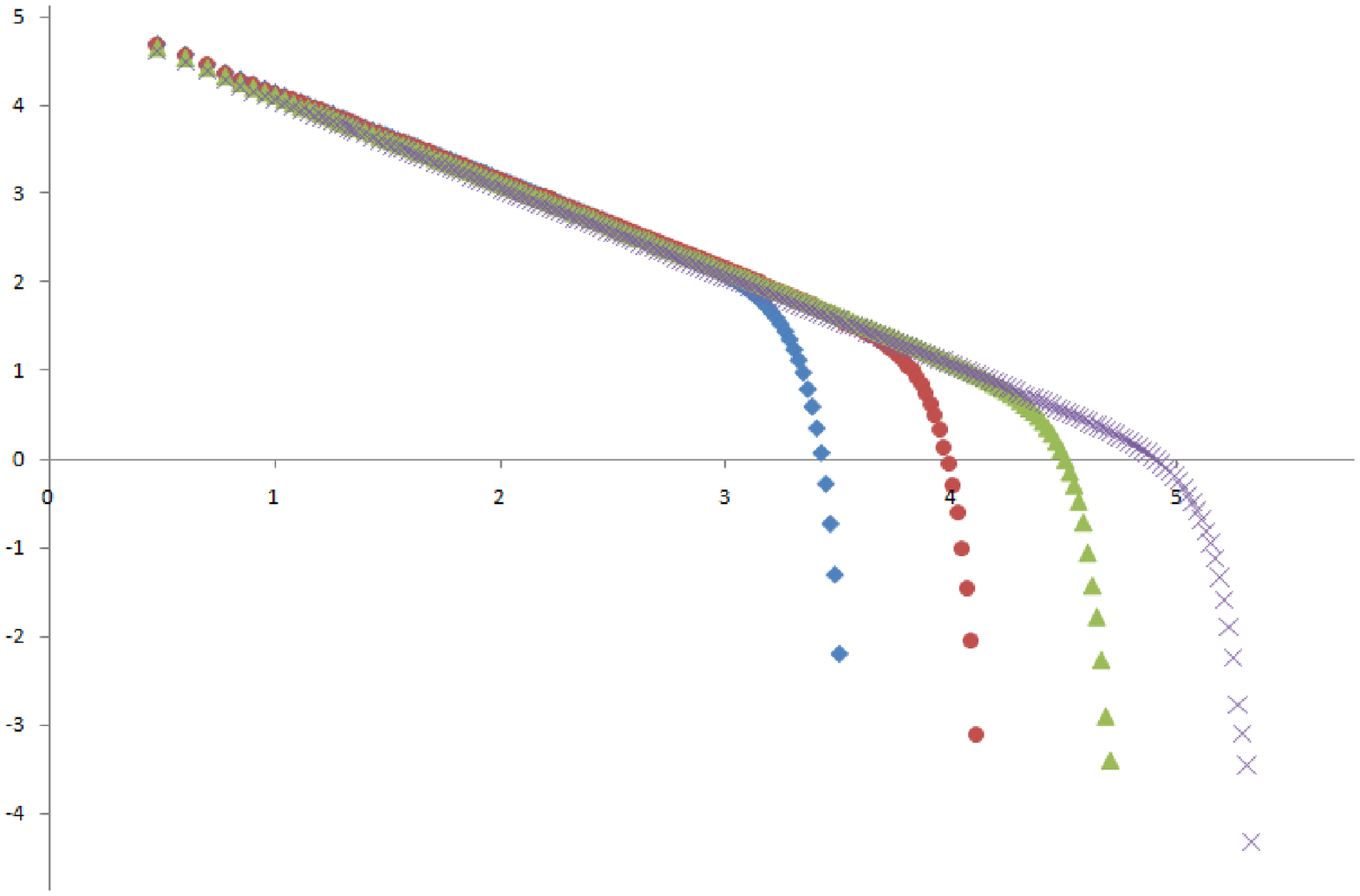}
\put(55,0){$\epsilon=0.1$}\put(220,0){$\epsilon=0.4$}\put(383,0){$\epsilon=0.8$}
\end{picture}
\caption{Wave size distribution for $\epsilon=0.1,\,0.4,\,0.8$ and
for lattice sizes $L=64,\,128,\,256,\,512$.}
\end{figure}
\begin{figure}[t]
\begin{picture}(800,180)(20,30)
\includegraphics{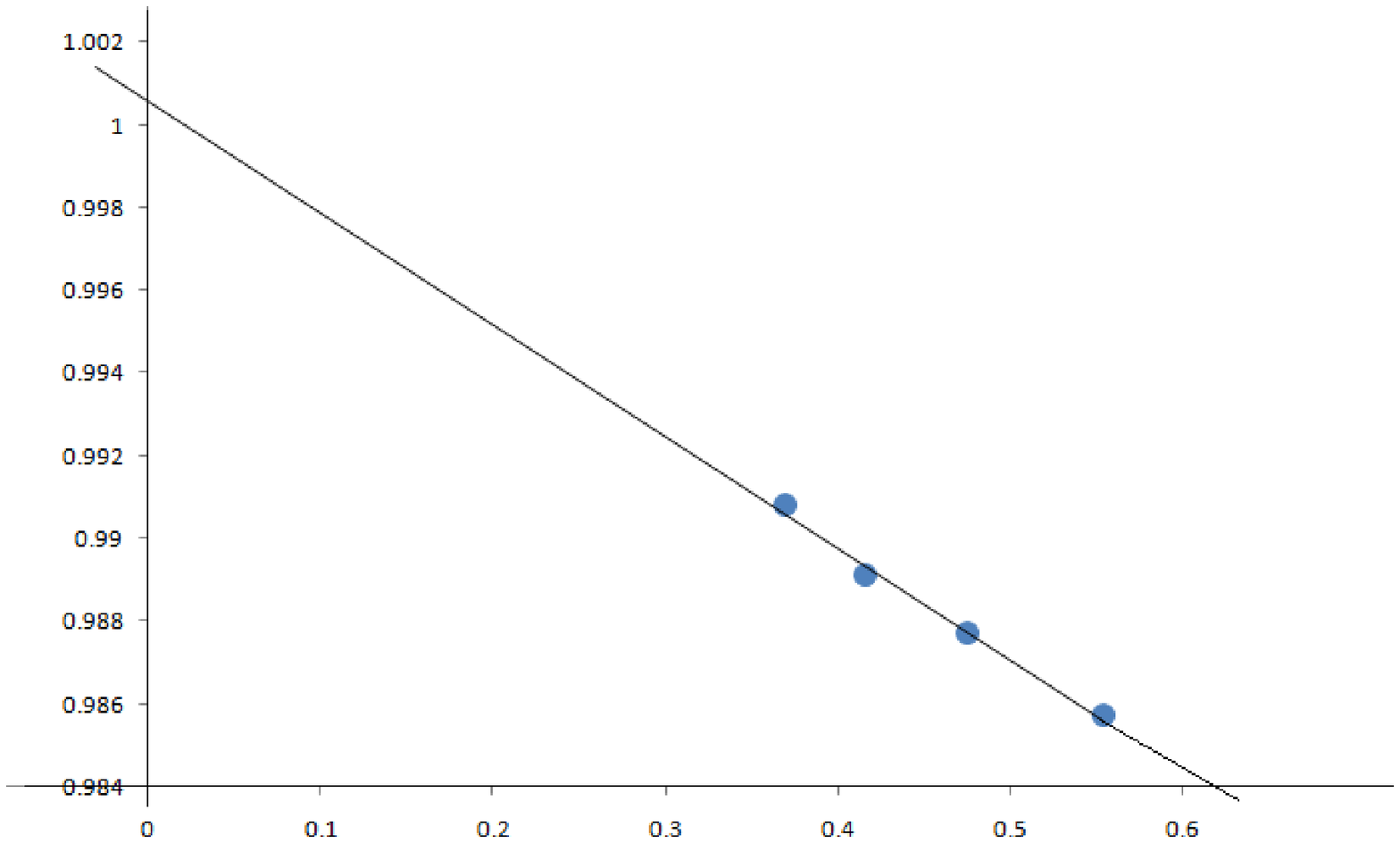}
 \includegraphics{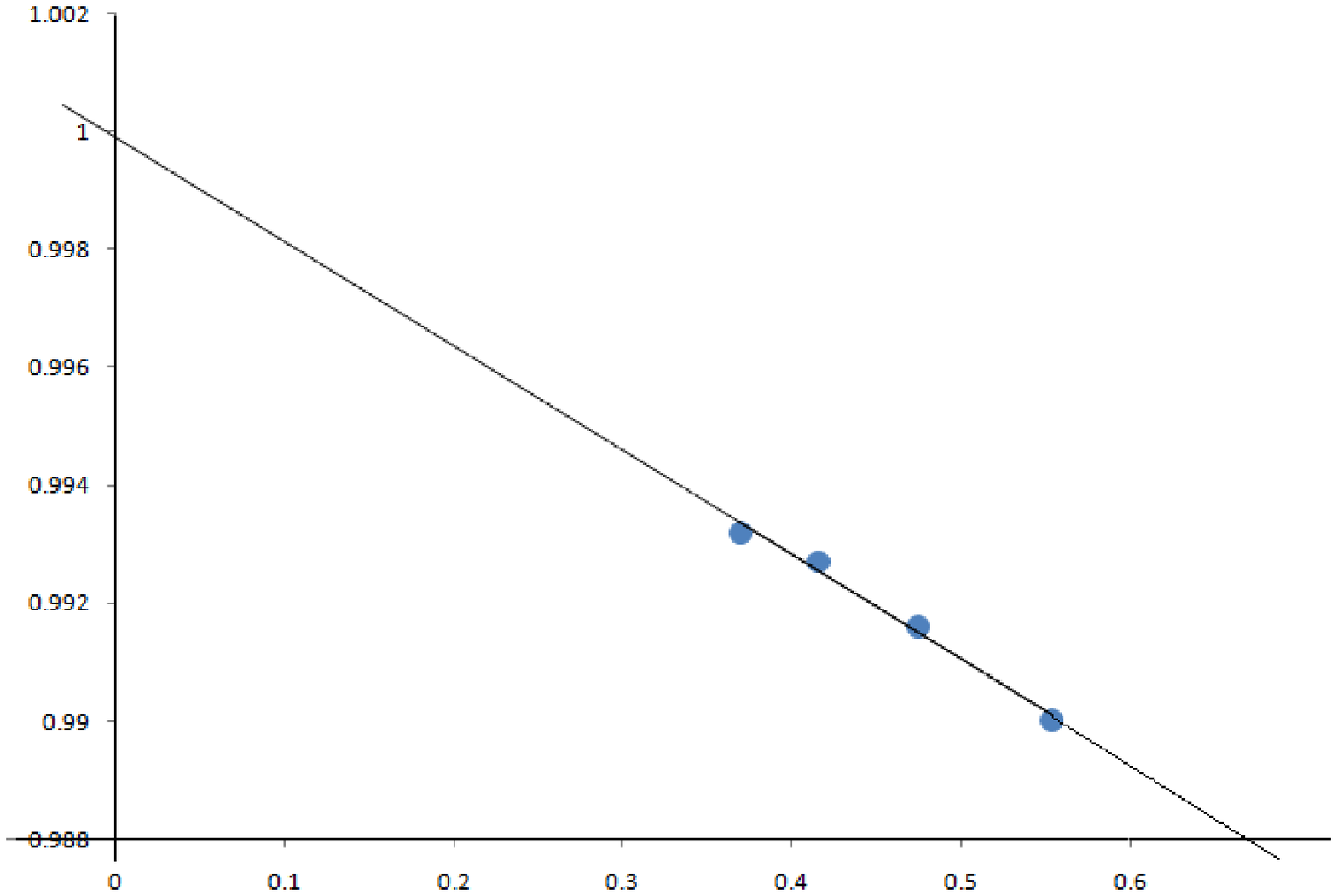} \includegraphics{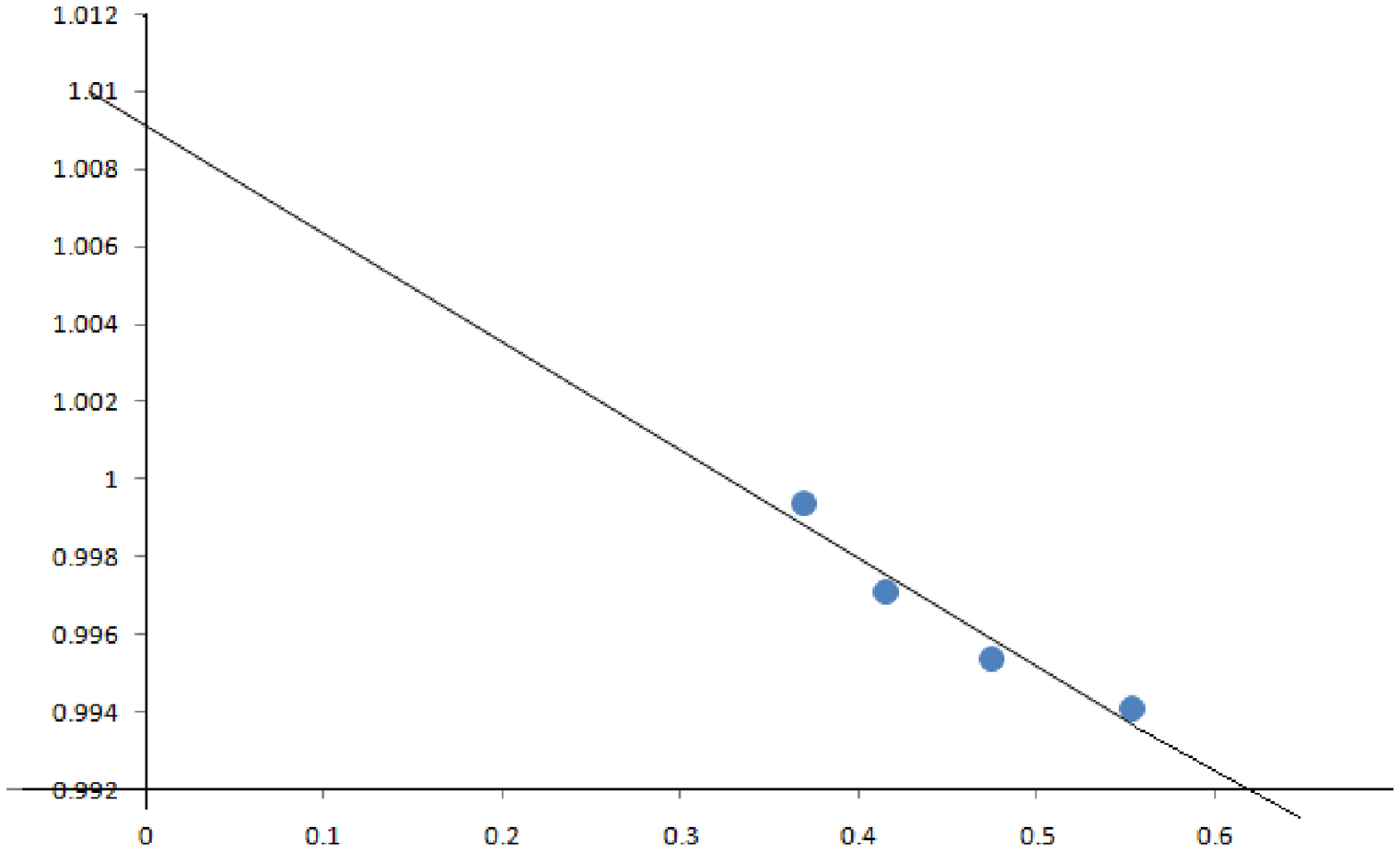}
\put(10,170){$\tau_s(L)$}\put(150,170){$\tau_s(L)$}
\put(90,70){$1/\log L$}\put(220,70){$1/\log L$}
\put(295,170){$\tau_s(L)$} \put(370,70){$1/\log L$}
\put(220,50){$\epsilon=0.4$} \put(370,50){$\epsilon=0.8$}
\put(90,50){$\epsilon=0.1$}
\end{picture}
\caption{The exponent $\tau(L)$ is a linear function of $1/\log
L$. The intersection with vertical axis gives $\tau_s(\infty)$}
\end{figure}
It is possible to reformulate the partition function or the number
of the spanning trees on the lattice in terms of fermionic path
integrals. We place a two-component Grassmannian variable
$\psi_{n}=(\psi_{1},\psi_{2})$ on each unit cell $n$ of the
lattice. In this representation, the action of the field theory is
written in the following form:
\begin{equation}\label{action}
S=\sum_{<n,\acute{n}>}\sum_{i,j=1}^{2}\psi^{\dag}_{i}(n)a_{ij}(n,\acute{n})\psi_{j}(\acute{n})
\end{equation}
where $a(n,\acute{n})$ are the adjacency matrices defined in (\ref{adjacency}). In the continuum
limit, this action is obtained to be:
\begin{equation}\label{actioncon}
S=\int dx
dy\sum_{\alpha,\beta=1}^{2}[4(-1)^{\alpha+\beta}\psi^{\dag}_{\alpha}(x)\psi_{\beta}(y)+2
\varepsilon^{\alpha\beta}(1-\epsilon)\partial_{x}\psi^{\dag}_{\alpha}(x)\psi_{\beta}(y)+2
\varepsilon^{\alpha\beta}\psi^{\dag}_{\alpha}(x)\partial_{y}\psi_{\beta}(y)]
\end{equation}
where $\varepsilon^{\alpha\beta}$ is the Levichivita
antisymmetric tensor. At the first sight it may look strange that we have an action that has only first derivative in it, in contrast with the $c=-2$ action that has second derivative terms. Even if we take the $\epsilon \rightarrow 0$ limit, it seems that the problem still exists. But if we look more closely, we will see that at least in the above limit one can write $\psi_2$ in terms of $\psi_1$ and its derivative and then the second-derivative terms emerge.

It may be argued that the above defined patterned system actually has a preferred direction; the zigzag paths join the down-left corner to up-right corner and not down-right to up-left. This is true, in fact the system has an elliptical anisotropy in the large scales and we know the elliptical anisotropy does not change the universality class \cite{asymmetry}. It is possible to introduce other patterns in a way that the system be symmetric in the large scales.    
Fig.5 shows such a pattern. In this model the thick lines
characterize bonds that carry $1+\epsilon$ amount of sand and the
thin lines carry $1-\epsilon$ amount of sand after a site
topples.
\begin{figure}[t]
\centering
\epsfig{file=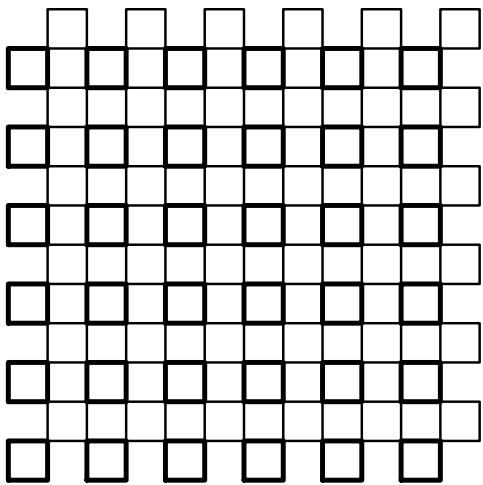, width=.4\linewidth} \caption{Order and
symmetric patterned ASM} \label{fig:}
\end{figure}
Following the standard procedure, the free energy of this system
is obtained:
\begin{eqnarray}\label{freeenergy3}
f&=&\frac{1}{16\pi^{2}}\int_{0}^{2\pi}d\theta\int_{0}^{2\pi}d\phi\ln\left[
132-136\epsilon^{2}+4\epsilon^{4}+2(1-\epsilon^{2})^{2}(\cos
2\theta+\cos 2\phi)\right.\nonumber\\&&\hspace{3cm}\left.-64(1-\epsilon^{2})(\cos
\phi+\cos\theta)-4(1-\epsilon^{2})^{2}(\cos(\theta+\phi)+\cos(\theta-\phi))\right]
\end{eqnarray}
which is again a smooth function and similar to the previous model and the self organized-criticality has the same universality class as of
the undirected sandpile model. 

Up to now, we have observed that the patterns that do not produce a preferred direction in large scales do not change the universality class. In the next section we will consider a quenched random anisotropy to see if the universality class is changed or not.

\section{Random directed continuous sandpile model }

In directed continuous sandpile model (DCSM) introduced in
\cite{asymmetry}, it has been assumed that after toppling of a site, $1+\epsilon$
amount of sand move to left (and up)  and $1-\epsilon$ amount of sand move
to right (and down); that is there exists a preferred direction for the
transportation of sands. In other words, the rotational
symmetry is broken in this model. In the continuum limit, It turns
out that the action of the theory assigned to the directed model
is the action of $c=-2$ conformal field theory perturbed by the
relevant scaling fields $\phi=-2\theta\partial\bar{\theta}$ and
$\bar{\phi}=-2\theta\bar{\partial}\bar{\theta}$. As these operators are relevant, they grow under renormalization and take the system to a new fixed point \cite{asymmetry}.

In DCSM, $\epsilon$ determines the strength of anisotropy and is in the interval $(-1,1)$. Positive $\epsilon$ means that the sand grains are pushed to the up-left corner and negative $\epsilon$ means that they are pushed to the down-right corner. In this model the value of $\epsilon$ is considered to be uniform through out the lattice. However, we may assume a statistical
distribution for $\epsilon$, such that it can take both positive and negative values on different sites. The assumption  that the mean value of $\epsilon$ vanish, means that there will be no
preferred direction statistically and the rotational symmetry will
be restored to the model. The question is if such a modification takes the system to a new universality class or not. The assumption of a weak randomness allows us to determine the critical behavior of the model based on the pertubative
renormalization group technique. 

In the continuous limit, the action of perturbed theory is given as:
\begin{eqnarray}\label{freeenergy3}
S=S_{0}+\int _{z} ~
\epsilon(z,\bar{z})(\phi(z,\bar{z})+\bar{\phi}(z,\bar{z}))
\end{eqnarray}
Where $S_{0}$ is the action of $c=-2$ logarithmic conformal field
theory. One can obtain the effective action using the replica
method; that is, we have to take average of 
$\epsilon$  on $N$ copies of the system and then find its limit when $N\rightarrow 0$. We assume that the $\epsilon(z)$ at different sites
are independent and have a Gaussian distribution  on each site with a standard deviation equal to $g_0$:
\begin{eqnarray}\label{freeenergy3}
\langle\epsilon(z_{1})\epsilon(z_{2})\rangle=g_{0}\delta(z_{1}-z_{2})
\end{eqnarray}
The effective
action then is expressed as:
\begin{eqnarray}\label{freeenergy3}
S=\sum_{a=1}^{N}S_{0,a}+ g_{0}\int _{z}\sum_{a\neq
b}^{N}(\phi_{a}(z,\bar{z})\phi_{b}(z,\bar{z})+\bar{\phi}_{a}(z,\bar{z})\bar{\phi}_{b}(z,\bar{z})
+\phi_{a}(z,\bar{z})\bar{\phi}_{b}(z,\bar{z}))
\end{eqnarray}
Although the coupling constants of the field operators
$\phi\phi$, $\bar{\phi}\bar{\phi}$ and $\phi\bar{\phi}$ are the same, as we will see, they have different RG equations. Therefore we distinguish the coupling constants of these field operators and rewrite them as
$g_{0\phi\phi}$, $g_{0\bar{\phi}\bar{\phi}}$ and
$g_{0\phi\bar{\phi}}$ respectively:
\begin{eqnarray}\label{freeenergy3}
\int _{z}\sum_{a\neq b}^{N}\left(g_{0\phi\phi}
\phi_{a}(z,\bar{z})\phi_{b}(z,\bar{z})+g_{0\bar{\phi}\bar{\phi}}\bar{\phi}_{a}(z,\bar{z})\bar{\phi}_{b}(z,\bar{z})
+g_{0\phi\bar{\phi}}\phi_{a}(z,\bar{z})\bar{\phi}_{b}(z,\bar{z})\right)\nonumber\\&&\hspace{-12.5cm}\equiv
\int _{z}\sum_{a\neq b}^{N}\Phi_{ab}(z)
\end{eqnarray}
where the second line is an abbreviation of the first line. It is easy to see that the coupling constants $g$ are dimensionless; that is, they are marginal. Therefore to see if they are marginally relevant or not, we have to expand the partition function to the second order of $g$. If it is marginally relevant we would like to see if it grows to infinity or will introduce a new fixed point. this means that we have to consider at least up to third order of coupling constants: 
\begin{eqnarray}\label{freeenergy3}
\int_{z}\sum_{a\neq
b}\Phi_{ab}(z)+\frac{1}{2!}\int_{z_{1},z_{2}}\sum_{a\neq
b}\Phi_{ab}(z_{1})\sum_{c\neq d}\Phi_{cd}(z_{2})
+\frac{1}{3!}\int_{z_{1},z_{2},z_{3}} \sum_{a\neq
b}\Phi_{ab}(z_{1})\sum_{c\neq d}\Phi_{cd}(z_{2})\sum_{e\neq f
}\Phi_{ef}(z_{3})\nonumber\\&&\hspace{-16cm}+\ldots
=g_{\phi\phi}\int_{z}\sum_{a\neq
b}\phi_{a}(z,\bar{z})\phi_{b}(z,\bar{z})+g_{\bar{\phi}\bar{\phi}}\int_{z}
\sum_{a\neq
b}\bar{\phi}_{a}(z,\bar{z})\bar{\phi}_{b}(z,\bar{z})+g_{\phi\bar{\phi}}\int_{z}
\sum_{a\neq b}\phi_{a}(z,\bar{z})\bar{\phi}_{b}(z,\bar{z})
\end{eqnarray}
To proceed, we have to know the contraction of fields in different possible
ways. The calculation is done using operator
product expansion (OPE) relations of the perturbing operators:
\begin{eqnarray}\label{freeenergy3}
\phi(z_{1},\bar{z}_{1})\phi(z_{2},\bar{z}_{2})&=&\frac{1}{(z_{1}-z_{2})^{2}}+\partial
\phi(z_{2},\bar{z}_{2})+2 T(z_{2},\bar{z}_{2})+\ldots\\
\bar{\phi}(z_{1},\bar{z}_{1})\bar{\phi}(z_{2},\bar{z}_{2})&=&\frac{1}{(\bar{z}_{1}-\bar{z}_{2})^{2}}+\partial
\bar{\phi}(z_{2},\bar{z}_{2})+2
\bar{T}(z_{2},\bar{z}_{2})+\ldots\\
\phi(z_{1},\bar{z}_{1})\bar{\phi}(z_{2},\bar{z}_{2})&=&\frac{1}{|z_{1}-z_{2}|^{2}}+\frac{\bar{\phi}(z_{1},\bar{z}_{1})}{z_{1}-z_{2}}-
\frac{\phi(z_{1},\bar{z}_{1})}{\bar{z}_{1}-\bar{z}_{2}}+\ldots
\end{eqnarray}
Where $T$ and $\bar{T}$ are the components of energy-momentum tensor. 

\begin{figure}[t]
\centering
\epsfig{file=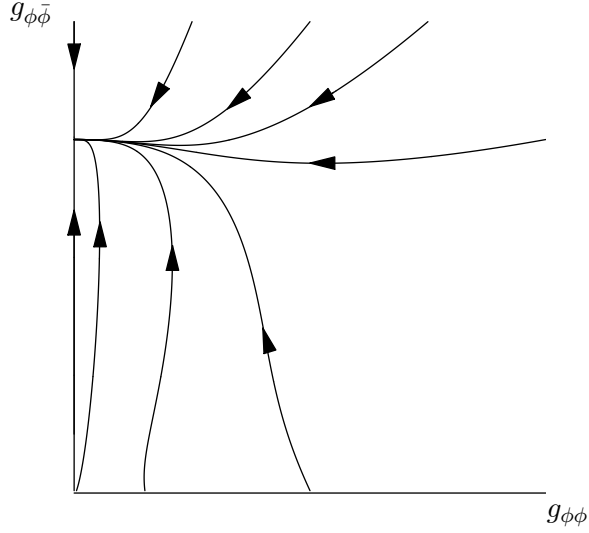, width=.4\linewidth}
\put(0,0){$g_{\phi\phi}$} 
\put(-203,190){$g_{\phi\bar{\phi}}$}\caption{The RG flow of the model with quenched randomness. } \label{fig6}
\end{figure}

At each order we contract all the fields using the above OPE relations and only keep a pair of $\phi$ or $\bar{\phi}$ fields. While doing the integrations we have to perform regularization. We do the regularization in cut-off scheme: we assume the
distance between any pair of integration variables is restricted
to be between $a$, the lattice constant, and $L$, size of the lattice. Up to the
third order, the renormalized couplings are obtained to be:
\begin{eqnarray}\label{freeenergy3}
g_{\phi \phi}=g_{0\phi \phi}+2\alpha(N-2)
g_{0\phi\phi}g_{0\phi\bar{\phi}}+2\alpha^{2}(N-2)[g_{0\phi
\phi}g_{0\phi\bar{\phi}}^{2}(5N-9)
+g_{0\bar{\phi}\bar{\phi}}g_{0\phi\phi}^{2}(3N-7)]
\end{eqnarray}
\begin{eqnarray}\label{freeenergy3}
g_{\bar{\phi}\bar{\phi}}=g_{0\bar{\phi}\bar{\phi}}+2\alpha(N-2)
g_{0\bar{\phi}\bar{\phi}}g_{0\phi\bar{\phi}}+2\alpha^{2}(N-2)[g_{0\bar{\phi}
\bar{\phi}}g_{0\phi\bar{\phi}}^{2}(5N-9)
+g_{0\phi\phi}g_{0\bar{\phi}\bar{\phi}}^{2}(3N-7)]
\end{eqnarray}
\begin{eqnarray}\label{freeenergy3}
g_{\phi\bar{\phi}}=g_{0\phi\bar{\phi}}+2
\alpha(N-3)(g_{0\phi\phi}g_{0\bar{\phi}\bar{\phi}}+g_{0\phi\bar{\phi}}^{2})
+8\alpha[g_{0\phi\bar{\phi}}^{3}\left((N-2)(N-1)+2(N-3)^{2}\right)+
\nonumber\\\hspace{6cm}g_{0
\phi\phi}g_{0\bar{\phi}\bar{\phi}}g_{0\phi\bar{\phi}}\left(3(N-2)(N-1)+2(N-3)^{2}\right)]
\end{eqnarray}
where $\alpha=4\pi\ln\frac{L}{a}$ and by the symmetry reasons,
$g_{\phi\phi}=g_{\bar{\phi}\bar{\phi}}$.  In the limit $N=0$, we
obtain the $\beta$-functions up to third order:
\begin{eqnarray}\label{freeenergy3}
\beta_{g_{\phi\phi}}&=&a \frac{\partial g_{\phi\phi}}{\partial
a}=16\pi
g_{\phi\phi}g_{\phi\bar{\phi}}-16\pi\alpha(9g_{\phi\phi}g_{\phi\bar{\phi}}^{2}
+7g_{\phi\phi}^{3})\\
\beta_{g_{\phi\bar{\phi}}}&=&a \frac{\partial
g_{\phi\bar{\phi}}}{\partial a}=24\pi(
g_{\phi\phi}^{2}+g_{\phi\bar{\phi}}^{2})-32\pi\alpha(5g_{\phi\bar{\phi}}^{3}
+6g_{\phi\phi}^{2}g_{\phi\bar{\phi}})
\end{eqnarray}
It is clear from above equations that these fields are marginally relevant, however the coefficients of the terms proportional $g^3$ are negative; hence the renormalization
flow takes the system to a fixed point at
$g_{\phi\phi}=g_{\bar{\phi}\bar{\phi}}=0$,
$g_{\phi\bar{\phi}}=\frac{3}{20\alpha}$(See Fig. \ref{fig6}). In the new random fixed
point, the rotational symmetry of the lattice restored so it is
expected that the system show critical behaviors different from
the deterministic directed model. 

We can compare our results with what Pan et al.\cite{pan} have found. In the patterned case, the outflow and inflow of the sand were balanced and we found that the universality class is not changed in such cases. On the other hand in the model with quenched randomness, there is not such a balance hence it is expected that the random fixed point belong to another universality class such as the universality class of the directed Manna sandpile model. We say it may correspond to Manna model because in this model there is randomness in the toppling rule, and we say it {\it may} be, because in Manna model the randomness is annealed but in our model it is quenched. 
\section{Conclusions}

In this paper we studied the critical behavior of the continuous
sandpile model with the some patterned anisotropies in toppling matrices.
Using the correspondence with the spanning trees, we obtained the
free energy function for theses models. Both theoretic analysis
and numerical simulations for the probability distribution of
waves indicate that the anisotropic models are in the same
universality
class of the continuous sandpile model.

Also we investigated analytically the effect of quenched
randomness on the critical behavior of continuous directed
sandpile model. Our calculations is based on the perturbed
renormalization conformal field theory and replica technique. Up to the third order
perturbation, we obtained the renormalization group equations for
the coupling constants of the perturbing fields. We showed that
the perturbing fields are relevant and take the system to the new
fixed point.

\end{document}